\documentclass[12pt,a4paper]{article}
\makeatletter
\def\fmslash{\@ifnextchar[{\fmsl@sh}{\fmsl@sh[0mu]}]}
\def\fmsl@sh[#1]#2{%
  \mathchoice
    {\@fmsl@sh\displaystyle{#1}{#2}}%
    {\@fmsl@sh\textstyle{#1}{#2}}%
    {\@fmsl@sh\scriptstyle{#1}{#2}}%
    {\@fmsl@sh\scriptscriptstyle{#1}{#2}}}
\def\@fmsl@sh#1#2#3{\m@th\ooalign{$\hfil#1\mkern#2/\hfil$\crcr$#1#3$}}
\makeatother
\global\arraycolsep=2pt 
\input{epsf}
\begin{document}
\thispagestyle{empty}
\rightline{CERN-TH/2001-090}
\rightline{TTP 01-08}
\rightline{hep-ph/0103331}
\rightline{March 2001}
\bigskip
\boldmath
\begin{center}
{\bf \Large
CP Asymmetries in $b \to (s/d)       $ Transitions \\[2mm]
as a Test of CKM CP Violation}
\end{center}
\unboldmath
\smallskip
\begin{center}
{\large{\bf Tobias Hurth$^{(a)}$ and Thomas Mannel$^{(a,b)}$ }}
\vspace*{2cm} \\
$^{(a)}${\sl CERN Theory Division, CH--1211 Geneva 23, Switzerland}\\

$^{(b)}$ {\sl Institut f\"{u}r Theoretische Teilchenphysik, \\
Universit\"{a}t Karlsruhe,  D--76128 Karlsruhe, Germany}\\
\end{center}

\begin{abstract}
\noindent
We point out that a simple test of the mechanism of CP
violation can be performed by a
measurement of the CP asymmetries in exclusive and inclusive
radiative rare decays. We show that the rate asymmetries
\mbox{
$\Delta \Gamma = \Gamma(B^- \to f \gamma) - \Gamma(B^+ \to \bar{f} \gamma)$}
for certain final states $f$
can be predicted in a theoretically clean way. Some implications for
$b \to s \ell^+ \ell^-$ decays are discussed. 
\end{abstract}
\newpage
\section{Introduction}
After the very successful start of the $B$ factories at
SLAC and KEK we may expect a large amount of data on decays
of $B$ mesons. 
Rare decays were first observed by the CLEO collaboration
\cite{Cleoexclusive,Cleoinclusive}; these measurements have been 
refined \cite{Cleoinclusive2} and confirmed by other experiments 
\cite{Alephinclusive,Belleinclusive}.
The theoretical prediction of the Standard Model (SM)
up to  next-to-leading logarithmic precision
for the total decay rate of the $b \rightarrow s \gamma$ mode 
is well in agreement with the experimental data \cite{NLL}.
The $b \rightarrow s \gamma$ mode already 
allows for theoretically clean and rather strong constraints on 
the parameter space of various extensions of the 
SM \cite{NLLbeyond,NLLbeyond2}.

Also, detailed measurements
of CP asymmetries in rare $B$ decays will be possible in the near 
future. Theoretical predictions for the {\it normalized} CP asymmetries of
the  inclusive channels (see~\cite{GreubAli,KaganNeubert,SoniWu}) within 
the Standard Model lead to 
\begin{equation}
\label{SMpredict} 
\nonumber 
\alpha_{CP}({b \rightarrow  s/d \, \gamma}) =
\frac{\Gamma(\bar B \rightarrow X_{s/d}\gamma)
     -\Gamma(B \rightarrow  X_{\bar s/\bar d}\gamma)}
     {\Gamma(\bar B \rightarrow  X_{s/d} \gamma)
     +\Gamma(B \rightarrow  X_{\bar s/\bar d}\gamma)}
\end{equation}
\begin{equation}                    
  \alpha_{CP}({b \rightarrow  s \gamma}) \approx 0.6 \%, \qquad 
  \alpha_{CP}({b \rightarrow d\gamma}) \approx  -16 \%
\label{SMnumbers}
\end{equation}
when the best-fit values for the CKM parameters \cite{CKMfit} 
are used. An analysis for the leptonic counterparts can be found in
\cite{Hiller}.
The normalized CP asymmetries may also be calculated for
exclusive decays; however, these results are model-dependent. An
example of such a calculation may be found in \cite{GSW}.

CLEO has already presented a measurement of the CP asymmetry in
inclusive $b \to s \gamma$ decays, yielding \cite{CleoCP}
\begin{equation}
\alpha_{CP}(b \rightarrow  s \gamma) = 
(-0.079 \pm 0.108 \pm 0.022) \cdot (1.0 \pm 0.030) \, , 
\end{equation}
which indicates that very large effects are already excluded.  
However, as we point out here, the decays of the form $b \to s \gamma$ and
$b \to d \gamma$ as well as their leptonic counterparts provide a
stringent test, if the CKM matrix is indeed the only source of
CP violation. We shall argue that the exclusive as well as the inclusive
decays may be used to perform a clean test of the CKM mechanism of CP
violation.

We start with the fact that the CP asymmetry for the sum of the partonic
processes $b \to (s+d) \gamma$ vanishes in the limit of $m_d = m_s = 0$. 
This was first observed by Soares \cite{Soares} in a partonic calculation
of these processes. Later Kagan and Neubert analysed the
CP asymmetry in inclusive $b \to s \gamma$ decays and mentioned this
as a side remark \cite{KaganNeubert}. This fact is still true when only the
weaker condition $m_d=m_s$ holds, which corresponds to the U-spin limit
at the quark level. 

We shall show in this note that one may use this fact for a stringent
test of the CKM mechanism of CP violation. Any CP violation in the
standard model has to be proportional to the determinant\footnote{
We assume here that the mass matrices for up and down quarks are
hermitean.}
\begin{eqnarray} \label{Jdet}
C &=& {\rm det} \left[ {\cal M}_U \, , \, {\cal M}_D \right]
\\ \nonumber 
&=& i \, J \,\,\, (m_u - m_c) (m_u - m_t) (m_c - m_t)
(m_d - m_s) (m_d - m_b) (m_s - m_b) 
\end{eqnarray}
where ${\cal M}_{U/D}$ are the mass matrices for the up and  down
quarks and 
\begin{equation} \label{Jarls}
J = {\rm Im}[V_{ub} V_{cb}^* V_{cs} V_{us}^*]
\end{equation}
is the Jarlskog parameter, which is a fourth-order quantity
and which is invariant under rephasing of the quarks fields. 
For a unitary CKM matrix there is exactly one such quantity,
and for that reason $J$ (or equivalently $C$) is the measure
of CP violation in the Standard Model.   However, comparing different
CP-violating processes involves also hadronic matrix elements, which in
general are hard to control, making a direct extraction of $C$ or $J$
in general difficult.

One way to gain information about the hadronic matrix elements is
to use symmetries. For the case at hand, we shall make use of the
U-spin, which is the $SU(2)$ subgroup of flavour $SU(3)$ relating
the $s$ and the $d$ quark. In fact U-spin symmetry is a well-known tool and
has already been used  
in the context of non-leptonic decays \cite{Fleischer:1999pa,GronauBCP4} and
general proofs exist for U-Spin relations among arbitrary heavy hadron decays 
\cite{Gronau:2000zy}.  In the present paper we shall apply this symmetry to
relate the decays $b \to s \gamma$ and $b \to d \gamma$. 

It is known from hadron spectroscopy
that the U-spin symmetry is violated, which at the parton level
originates from the different masses of the down and the strange
quark. Furthermore, if the down and the strange quark were degenerate, 
the Standard Model would be CP-conserving, as can be seen from
(\ref{Jdet}).

However, we shall make use of this symmetry only
with respect to the strong interactions; although the down and
strange quark masses are different, we shall consider (hadronic)
final states
with masses well above the down and strange quark (current) masses,
and thus the U-spin limit is still useful.

In section~\ref{secexc} we discuss the U-spin relations
for hadronic matrix elements relevant to the exclusive decays;
in section~\ref{secinc} we study inclusive processes. 
In section~\ref{beyond} we conclude with a short discussion of possible non-standard model
scenarios.

\section{U-Spin Relations: Exclusive Decays} \label{secexc}
The effective Hamiltonean mediating rare radiative or rare semileptonic
transitions  from $b \to q \gamma$ ($q = d,s$) can be decomposed
into the pieces with different weak phases
\begin{equation} \label{heff}
H_{eff} = \lambda_u^{(q)} A + \lambda_c^{(q)} B + \lambda_t^{(q)} C
        = \lambda_u^{(q)} (A-C) + \lambda_c^{(q)} (B-C),
\end{equation}
where
\begin{equation}
\lambda_u^{(q)} = V_{ub} V_{uq}^* \qquad
\lambda_c^{(q)} = V_{cb} V_{cq}^* \qquad
\lambda_t^{(q)} = V_{tb} V_{tq}^*
\end{equation}
and where we have used the unitarity relation
$\lambda_u^{(q)} + \lambda_c^{(q)}+ \lambda_t^{(q)} = 0$ 
to eliminate $\lambda_c^{(q)}$.

As far as the strong interaction U-spin is concerned, the effective
Hamiltonian transforms as a doublet under this symmetry, i.e.
$A$, $B$ and $C$ are U-spin doublets.
The charged $B$ mesons contain neither the $d$ nor the $s$ flavour
and are thus U-spin singlets. The neutral $B$ mesons
$(B_d, \,\,B_s)$ form a doublet under U-spin, which makes their case
more complicated.

The final states for the  exclusive processes we can consider
are the $\pi$, $K$, $\rho$ and $K^*$ states. Here again
the case of charged states is simple: The
$(\pi^- \,\, K^-)$ and the $(\rho^- \,\, K^{*-})$ form
two U-spin doublets. For the neutral mesons we can form
a U-spin singlet and a U-spin triplet. For the
vector mesons (assuming that the $\phi$ is a pure
$s\bar{s}$ state) the singlet is a combination of $\phi$, $\rho$ and
$\omega$: $
{1}/{\sqrt{2}} [ |\phi\rangle
+ \frac{1}{2}(|\omega \rangle - |\rho \rangle) ]$
while for the triplet we have 
the $|K^0\rangle$ and $|\overline{K}^0\rangle$
as the $\pm 1$ component, and  
${1}/{\sqrt{2}} [|\phi\rangle
- \frac{1}{2}(|\omega \rangle - |\rho \rangle) ]$
for the $0$ component.

As mentioned above, the U-spin is clearly  broken by the different
masses $m_d$, $m_s$ of the down and the strange quark. On the
other hand, both $m_d$ and $m_s$ are small with respect to the masses
of any hadron, except for the octet of light pseudoscalars, the
masses of which vanish in the chiral limit of QCD and thus 
are presumably more sensitive to the (current quark) masses of
the $s$ and  $d$ quarks. 

To this end, when talking about exclusive final states,
we shall mainly consider the vector mesons;
these have masses much larger than the (current quark) masses
of any of the light quarks. Thus we expect, for the ground state 
vector mesons, the U-spin symmetry to be quite accurate inspite of the
non-degeneracy of $m_d$ and $m_s$. A measure for the breaking of
U-spin symmetry is certainly the relative mass difference between
the $\rho$ and the $K^*$,  which is of the order of fifteen percent.  

For the decay of neutral mesons such as $B_d \to \rho^0 \gamma$
or $B_d \to \pi_0 \ell^+ \ell^-$, the U-spin relations 
involve two reduced matrix elements corresponding to the two
possibilities to couple the U-spins: the doublet of $B$ mesons
can be coupled with the (doublet) Hamltonian either to a singlet or
a triplet. The two matrix elements can in principle be disentangled 
by measuring all the decays of neutral $B$ mesons, including
processes like $B_s \to \phi \gamma$. Since this will
not be possible in the near future, we shall concentrate
on what can be done at the $B$ factories.

Thus the charged modes are more promising since they do not involve
any $B_s$ decay. Let us first consider the radiative decay
$B^\pm \to V^\pm \gamma$, where $V = \rho$ or $K^*$. The rate
for these decays may be written as
\begin{equation} \label{rate}
\Gamma (B^- \to V^- \gamma) = | \lambda_u M_u + \lambda_c M_c |^2,
\end{equation}
where we have suppressed the superscript $(q)$ for simplicity,
and $M_u$ is the matrix element of $A-C$ and $M_c$ is the one of
$B-C$.

The charge-conjugate process is
\begin{equation}
\Gamma (B^+ \to V^+ \gamma) = | \lambda_u^* M_u + \lambda_c^* M_c |^2,
\end{equation}
which yields for the rate difference
\begin{eqnarray} \label{ratediff}
\Delta \Gamma (B^- \to V^- \gamma) &=&
\Gamma (B^- \to V^- \gamma) - \Gamma (B^+ \to V^+ \gamma) \\ \nonumber
&=& - 4 \, {\rm Im} (M_u M_c^*) \, {\rm Im} (\lambda_u \lambda_c^*).
\end{eqnarray}

This is a standard expression showing that CP violation is
indeed proportional to $J$, since
${\rm Im} (\lambda_u \lambda_c^*) = \pm J$.

The hadronic matrix elements for $V=\rho$ and $V=K^*$ 
are related by U-spin. Comparing
\mbox{$B^\pm \to K^{*\pm} \gamma$} with \mbox{$B^\pm \to \rho^\pm \gamma$},
one finds (up to U-spin-breaking effects):
\begin{equation}
M_u^{(K^*)} = M_u^{(\rho)} = M_u \qquad
M_c^{(K^*)} = M_c^{(\rho)} = M_c,
\end{equation}
which means that
\begin{eqnarray}
\Delta \Gamma (B^- \to K^{*-} \gamma) &=&
- 4 \, {\rm Im} \, (M_u M_c^*) \, {\rm Im} \,
(\lambda_u^{(s)} \lambda_c^{(s)*}) \\
\Delta \Gamma (B^- \to \rho^- \gamma) &=&
- 4 \, {\rm Im} \, (M_u M_c^*) \, {\rm Im} \,
(\lambda_u^{(d)} \lambda_c^{(d)*}).
\end{eqnarray}
Using unitarity of the CKM matrix one can show easily that
\begin{equation}
J = {\rm Im} (\lambda_u^{(s)} \lambda_c^{(s)*})
= - {\rm Im} (\lambda_u^{(d)} \lambda_c^{(d)*})
\end{equation}
and thus one finds that the rate asymmetries ({\it not} 
the CP asymmetries,
which are the rate asymmetries normalized to the sum of the rates)
satisfy - in the U-spin limit for the hadronic matrix elements - 
the relation
\begin{equation} \label{resexc}
\Delta \Gamma (B^- \to K^{*-} \gamma) = -
\Delta \Gamma (B^- \to \rho^- \gamma).
\end{equation}
This result implicitly may be found in the literature,
but to our knowledge nobody has yet considered its implications
in detail.

Relation (\ref{resexc}) provides us with a relatively clean test of the
CKM mechanism of CP violation. The fact that $J$ is the only CP-violating
parameter in the Standard Model is deeply related to the unitarity
of the CKM matrix and to the fact that there are only three families.
Any non-SM scenario has generically more sources of
CP violation, which in general would disturb (\ref{resexc}), since
there will be other weak phases besides the CKM phase.

Clearly the main uncertainty in (\ref{resexc}) is U-spin breaking,
which means that (\ref{resexc}) is only useful with some estimate of
this breaking. We shall prceed in a similar way as in \cite{Fleischer:1999pa} 
and write
\begin{eqnarray} 
&& \Delta \Gamma (B^- \to K^{*-} \gamma) +
\Delta \Gamma (B^- \to \rho^- \gamma) \\ \nonumber 
&& \qquad \qquad 
= - 4 \,  J \,   {\rm Im} \left(M_u^{(K^*)} (M_c^{(K^*)})^*
                    - M_u^{(\rho)} (M_c^{(\rho)})^* \right)
= b_{exc} \Delta_{exc}
\end{eqnarray}
where the right hand side is written as a product of a ``relative 
U-spin breaking'' $b_{exc}$ and a ``typical size'' $\Delta_{exc}$ 
of the CP violating rate difference. Explicitely we have 
\begin{equation}
b_{exc} =  \frac{{\rm Im} \left(M_u^{(K^*)} (M_c^{(K^*)})^*
                    - M_u^{(\rho)} (M_c^{(\rho)})^* \right)}
 {\frac{1}{2}{\rm Im} \left(M_u^{(K^*)} (M_c^{(K^*)})^*
                    + M_u^{(\rho)} (M_c^{(\rho)})^* \right)}
\end{equation}
and
\begin{equation}
\Delta_{exc} =  - 2 \, J \, {\rm Im} \left(M_u^{(K^*)} (M_c^{(K^*)})^*
                      + M_u^{(\rho)} (M_c^{(\rho)})^* \right) 
\end{equation}
which is half of the difference of the two rate asymmetries.

The advantage to write the right hand side as $b_{exc} \Delta_{exc}$
is that, although we know neither $b_{exc}$ nor  $\Delta_{exc}$ 
precisely, we still can estimate it. 
$\Delta_{exc}$ can only be computed in a model, 
but the relative breaking $b_{exc}$ of U-spin can be estimated
e.g. from spectroscopy. This leads us to 
\begin{equation}
|b_{exc}| =  \frac{M_{K^*} - m_\rho}{\frac{1}{2}(M_{K^*} + m_\rho)}
= 14 \% 
\end{equation}
which takes into account our ignorance about the sign of $b_{exc}$.
Certainly also other estimates are possible, such as a comparison of
$f_\rho$ and $f_{K^*}$\footnote{There are various definitions for these 
quantities in the literature; we choose to define them as 
$\langle V(p,\epsilon)| j_\mu | 0 \rangle = f_V m_V \epsilon_\mu$.};  both
quantities can be determined from $\tau$  decays and one finds in this case a
very small U-spin breaking. We shall use the more conservative value for
$b_{exc}$, which is also compatible with sum rule calculations of form factors 
(see  \cite{Ali:1994vd}). 

For $\Delta_{exc}$ we use the model result  from \cite{GSW} and get 
\begin{equation}
\Delta_{exc}  = 2.5 \cdot 10^{-7}\,\,  \Gamma_B
\end{equation}
which leads us finally to our standard-model prediction for the
difference of branching ratios
\begin{equation} \label{resexc1}
|\Delta Br (B^- \to K^{*-} \gamma) +
\Delta Br (B^- \to \rho^- \gamma)| \sim 4 \cdot 10^{-8}
\end{equation}
Note that we can neither give a precise value nor the sign of the U-spin
breaking,  since the right hand side is model-dependent.  Still
(\ref{resexc1}) is of some use, since a value
significantly above this estimate would be a strong hint to non-CKM
contributions to CP violation.  

\boldmath
\section{Inclusive Decays and $b \to s \ell^+ \ell^-$ } \label{secinc}
\unboldmath
One may use similar arguments for the case of inclusive decays. For the
inclusive radiative rare decays of charged $B$ meson decays, exactly the same
arguments hold for any arbitrary final state, and thus we have for the rate
asymmetries the relation
\begin{equation} \label{resinc}
\Delta \Gamma (B^- \to X_s \gamma) = -
\Delta \Gamma (B^- \to X_d \gamma).
\end{equation}
Concerning the validity of U-spin, similar arguments hold. Since the
lowest state in the radiative decay is a vector meson\footnote{There
could also be a non-resonant contribution from $K \pi$ and
$\pi \pi$ states with a mass lower than that of the corresponding
vector meson, but this is known to be small.}
the invariant masses of the final states are large with respect to
$m_d$ and $m_s$, so we expect the U-spin to be a fairly good symmetry,
very likely even better than for the exclusive channels (see below).

Going one step further one may employ the $1/m_b$ expansion for the
inclusive process. To leading order the inclusive decay rate is
the free $b$-quark decay. In particular, there is no sensitivity to
the spectator quark and thus we may generalize (\ref{resinc}) and
include also neutral $B$ mesons
\begin{equation} \label{resincg}
\Delta \Gamma (B \to X_s \gamma) = -
\Delta \Gamma (B \to X_d \gamma)
\end{equation}

Furthermore, in this framework one relies on
parton-hadron duality (besides in the long-distance contribution
from up-quark loops which is found to be rather small \cite{LDUP}). 
So one can actually compute the breaking 
of U-Spin by keeping a non-vanishing strange quark mass. 
However, it is a formidable task to do this for the CP asymmetries
and it has not yet been done. Still it is clear that the relevant
parameter in the dominating contribution is $m_s^2/m_b^2$, which is
very small. 

Thus, we again parametrize the size of U-spin breaking in the same 
way as for the exclusive decays. We write 
\begin{equation} \label{resincg1}
\Delta \Gamma (B \to X_s \gamma) +
\Delta \Gamma (B \to X_d \gamma) = b_{inc} \Delta_{inc} 
\end{equation}
where now the typical size of $b_{inc}$ can be roughly 
estimated to be of the order
$|b_{inc}| \sim m_s^2/m_b^2 \sim 5 \cdot 10^{-4}$ . 
$|\Delta_{inc}|$  is again the average of the moduli of the two CP rate
asymmetries. These have been calculated (for vanishing strange quark mass)
e.g. in  \cite{GreubAli} and thus we arrive at 
\begin{equation} \label{resinc3}
| \Delta Br (B \to X_s \gamma) +
\Delta Br (B \to X_d \gamma) | \sim 1 \cdot 10^{-9} 
\end{equation}
Again, any measured value in significant deviation of (\ref{resinc3}) 
would be an indication of new sources of CP violation. Although we give
only an estimate here, we point out again that in the inclusive mode the
right-hand side in (\ref{resinc3})  can be computed in a
model-independent way with the help  of the heavy mass expansion. 

Going beyond leading order in the $1/m_b$ involves corrections of
order $\lambda_1 /m_b^2$ and $\lambda_2 / m_b^2$ which are small and
cancel in the sum of the rate asymmetries - 
in the limit of U-spin symmetry. 
Corrections of order $1/m_b^3$
involve also contributions (for example annihilation topologies), which distinguish between the charged
and the neutral $B$ mesons. These contributions 
are actually suppressed relative to the leading order by 
$\alpha_s(m_b)/m_b^3$ and are thus small; furthermore, the final
states originating from annihilation carry neither strangeness nor a
down quantum number. Strictly speaking these diagrams do not
contribute to $X_s$ or $X_d$ respectively, but it is a question of
the experimental set up how strongly states without $s$ or $d$ quantum
numbers can be discriminated. 

Finally, we make a few remarks on transitions of the form
$b \to q \ell^+ \ell^-$, $q=s,d$. Since the CKM structure of these
decays is the same, one may use eqs. (\ref{heff} -- \ref{ratediff})
in the same way, only with different non-hadronic contributions
to the final states. Thus, we arrive at
the same conclusions for the CP-violating rate differences. Of course,
the use of U-spin may be questionable for the light pseudoscalars, but
for the vector mesons the above arguments apply.

Moreover, since $B^\pm \to V^\pm \ell^+ \ell^-$, $V = K^{*\pm}, \rho^\pm$
is a three-body decay, one may perform additional tests. Although this is
not related to the CKM phases, one would end up with the prediction that
the forward--backward {\em rate} asymmetry would be the same for
$K^*$ and $\rho$. This prediction could be used as an additional 
cross-check for the validity of our assumption of U-spin symmetry 
in the  $K^{*\pm}$ -- $\rho^\pm$ system and maybe even access U-spin
breaking in a model-independent way also for the exclusive modes.

\section{``New Physics''}~\label{beyond}
Clearly the pattern of CP violation is very peculiar in the
SM. In other words, any new physics contribution
is likely to have additional sources for CP violation, i.e.\
additional weak phases violating (\ref{resexc1}) or 
(\ref{resinc3}). Let us conclude with a short look at scenarios
beyond the SM.

Although it has its well-known flavour problem, 
supersymmetry is given priority as a candidate for physics
beyond the SM. Supersymmetric predictions for the CP asymmetries in 
$b \rightarrow {s/d} \gamma$ depend strongly on what is
assumed for the supersymmetry-breaking sector and are thus  
a rather model-dependent issue. 
The minimal supergravity model cannot account for large 
CP asymmetries beyond $2\%$ because of the constraints coming 
from the electron and neutron electric dipole moments
\cite{Goto}. However, more general models 
allow for larger asymmetries of the order of $10 \%$ 
or even larger \cite{Aoki,KaganNeubert}.
Recent studies of the $b \to d \gamma$ rate asymmetry  
in specific models led to asymmetries between $-40 \% $ and $+40\%$
\cite{Recksiegel} or  $-45 \%$ and $+ 21 \%$ \cite{Asatrian}.

In general, CP asymmetries may lead to clean 
evidence for  new physics by a significant deviation from the SM
prediction. 
From (\ref{SMnumbers}) it is obvious  that a large CP asymmetry 
in the 
$b \rightarrow s \gamma$ channel or a positive CP asymmetry in the 
inclusive $b \rightarrow d \gamma$ channel would be a clear signal
for new physics. However, if indeed a CP asymmetry in conflict with
the SM is observed, this will only be an indirect hint to physics
beyond the SM, and it will be difficult to identify
the new structures in detail with only the information from
$B$ physics. Our test could  help to discriminate between the
different possibilities; 
It provides a definite test if generic new CP phases are
present or not since  it is rather unlikely that 
relations like (\ref{resexc1}) and (\ref{resinc3}) 
hold if new sources of CP violation are active.

\section{Conclusion}
A clean test of the SM pattern of CP violation is clearly one of the
main topics of on-going experiments, in particular at the $B$
factories. Being a direct CP asymmetry, the asymmetries in
$b \to (s/d) \gamma$ can be measured without the information on the
proper decay time.

Despite their smaller branching ratios, 
the exclusive channels are easier to identify, and we expect an
experimental test of (\ref{resexc}) in the near future. However,
for these decays the issue of U-spin breaking introduces a model
dependence. Measuring the corresponding forward--backward 
rate asymmetry may allow for a model-independent estimation
of the U-spin breaking in this channel.

The inclusive relations are theoretically cleaner, since one can
actually compute the U-spin breaking in this case. 
Parametrically the U-spin breaking in the leading contribution will 
be of order $m_s^2 / m_b^2$ and thus
very small. From the experimental side, the inclusive mode
is more difficult, since one has to make sure that the final state has
the strange or the down quantum numbers. However, contributions
which do not satify this criterion are suppressed by at least $1/m_b^3$.
and thus do not contaminate the inclusive measurement
$B \to (X_s + X_d) \gamma$.  

\section*{Acknowledgements}
We thank Gerhard Buchalla, Andrew Ackeroyd and Stefan Recksiegel
for useful discussions, and Patricia Ball for useful comments 
on the manuscript. TM is supported by the DFG 
Forschergruppe
``Quantenfeldtheorie, Computeralgebra und Monte Carlo Simulationen'' 
and from the Ministerium f\"ur Bildung und Forschung bmb+f.

\end{document}